\documentclass[twocolumn,amsmath,amssymb,superscriptaddress,notitlepage]{revtex4-1}
\usepackage{graphics}
\usepackage{bm}
\usepackage{dcolumn}
\usepackage{natbib}
\usepackage{epstopdf}
\usepackage{float}
\usepackage{graphicx}
\usepackage{epsfig}
\usepackage[pdfstartview=FitH]{hyperref}
\usepackage{color}
\usepackage{appendix}
\usepackage{ulem}

\newcommand{\rr}{{\bf r}}

\begin{document}
\title{Fractional quantum Hall effect of Bose-Fermi mixtures}
\author{Tian-Sheng Zeng}
\affiliation{Department of Physics, School of Physical Science and Technology, Xiamen University, Xiamen 361005, China}
\date{\today}
\begin{abstract}
Multicomponent quantum Hall effect, under the interplay between intercomponent and intracomponent correlations,
leads us to new emergent topological orders. Here, we report the theoretical discovery of fractional quantum hall effect of strongly correlated Bose-Fermi mixtures classified by the $\mathbf{K}=\begin{pmatrix}
m & 1\\
1 & n\\
\end{pmatrix}$ matrix (even $m$ for boson and odd $n$ for fermion), using topological flat band models. Utilizing the state-of-the-art exact diagonalization and density-matrix renormalization group methods, we build up the topological characterization based on three inherent aspects: (i) topological $(mn-1)$-fold ground-state degeneracy equivalent to the determinant of the $\mathbf{K}$ matrix, (ii) fractionally quantized topological Chern number matrix equivalent to the inverse of the $\mathbf{K}$ matrix, and (iii) two parallel-propagating chiral edge branches with level counting $1,2,5,10$ consistent with the conformal field theory description.
\end{abstract}
\maketitle

\textit{Introduction.---}
The discovery of the fractional quantum Hall (FQH) effect affords us a new paradigm of topologically ordered phases~\cite{Wen2017}, characterized by fractionalization and long-range entanglement. Beyond one-component systems, multicomponent quantum Hall effects endow us with the emergence of new kind of topological order, which expands the taxonomy of topological phases of matter, such as Halperin's two-component quantum Hall effect~\cite{Halperin1983} and quantum Hall ferromagnet~\cite{Wen1992} experimentally manifest in double-layer electron systems~\cite{Eisenstein1992,Suen1992}. Recent experiments have even revealed the emergence of an abundance of four-component FQH effects in graphene~\cite{Bolotin2009,Dean2011} when electronic multiple internal degrees of freedom are included. More spectacularly, new types of correlated many-body topological states in two parallel graphene layers have also been observed~\cite{Liu2019,Li2019}, which may be explained with next-level composite fermions.
Similarly, without available electronic analogs, two-component bosons are shown to bring up new competing quantum Hall structures including Halperin $(221)$ FQH effect~\cite{Grass2012,Furukawa2012,Wu2013,Grass2014} and bosonic integer quantum Hall effect~\cite{Senthil2013,Furukawa2013,Regnault2013,Wu2013,Grass2014} based on cold atomic neutral systems.

To date however, rare clear-cut examples concerning topologically ordered phases of Bose-Fermi mixtures which possess diverse quantum particle statistics, have been conclusively sought for. Most prior theoretical and experimental efforts were focused on topologically trivial symmetry-broken phases~\cite{Lewenstein2004,Salomon2014}. It remains an outstanding elusive question whether new kinds of FQH effect of Bose-Fermi mixtures can be produced or not. Theoretically for any Abelian multicomponent quantum Hall effect, its intrinsic topological structure can be captured by the $\mathbf{K}$ matrix within the framework of the Chern-Simons gauge-field theory~\cite{Wen1992a,Wen1992b,Blok1990a,Blok1990b,Blok1991}, where the diagonal and off-diagonal elements of the $\mathbf{K}$ matrix are successfully derived from the inverse of the Chern number matrix~\cite{Zeng2017,Zeng2018,Zeng2019,Zeng2020}. Provided the Abelian topological order for two-component Bose-Fermi mixtures, one may postulate that it should be intactly classified by the $\mathbf{K}=\begin{pmatrix}
m & l\\
l & n\\
\end{pmatrix}$ matrix,
at boson fillings $\nu_b=(n-l)/(mn-l^2)$ and fermion filling $\nu_f=(m-l)/(mn-l^2)$, where the diagonal element $m$ is even for bosons, $n$ is odd for fermions and the symmetric off-diagonal element $l\in \mathbb{Z}$. Nevertheless, how to realize such a topological order of Bose-Fermi mixtures has long remained an unfulfilled challenging area to be uncovered with much scope for theoretical surprises,
aside from the practical experimental interest. Most recently, several chiral short-range-entangled topological states without fractionalized excitations in Bose-Fermi mixtures based on two-dimensional synthetic gauge fields, were also numerically proposed with specific equal matrix elements $n=l=1$~\cite{Wu2019}. Instead, we focus on the identification of long-range-entangled fractionalized topological phases with unequal matrix elements $n\neq l$ and nontrivial topological degeneracy of the $\mathbf{K}$ matrix.

In this Letter, we investigate the topological order of Bose-Fermi mixtures in topological flat band models which have sparked a fertile vein to explore emerging strongly correlated Chern insulators driven by interactions~\cite{Chen2020,Ge2020,Wu2020,Nuckolls2020,Choi2020}. Both early theoretical studies~\cite{Sun2011,Neupert2011,Sheng2011,Wang2011,Regnault2011} and experimental observations~\cite{Spanton2017} identified that fractionalized topologically ordered phases can emerge at partial fillings of topological bands when strong interaction is introduced. In cold atom experiments, the potential to create iconic topological models for either bosons or fermions, such as Haldane-honeycomb and Harper-Hofstadter models~\cite{Jotzu2014,Aidelsburger2013,Miyake2013,Mancini2015,Stuhl2015}, laying the groundwork to explore new aspects of FQH physics. Indeed, experimental Bose-Fermi mixtures in optical lattice exhibit strong correlation effect arising from intercomponent interaction~\cite{Best2009,Sugawa2011}.

\textit{Model Hamiltonian.---}
Here, we will numerically address the emergence of FQH effect of Bose-Fermi mixtures with strong interactions in topological flat bands through state-of-the-art density-matrix renormalization group (DMRG) and exact diagonalization (ED) simulations. Two-component Bose-Fermi mixtures are strongly coupled via Bose-Bose, Fermi-Fermi and Bose-Fermi repulsions in two prototypical topological lattice models~\cite{sm}, such as the $\pi$-flux checkerboard (CB) lattice,
\begin{align}
  H_{CB}=&\!\sum_{\sigma=b,f}\!\Big[-t\!\!\sum_{\langle\rr,\rr'\rangle}\!e^{i\phi_{\rr'\rr}}c_{\rr',\sigma}^{\dag}c_{\rr,\sigma}
  -\!\!\!\!\sum_{\langle\langle\rr,\rr'\rangle\rangle}\!\!\!t_{\rr,\rr'}'c_{\rr',\sigma}^{\dag}c_{\rr,\sigma}\nonumber\\
  &-t''\!\!\!\sum_{\langle\langle\langle\rr,\rr'\rangle\rangle\rangle}\!\!\!\! c_{\rr',\sigma}^{\dag}c_{\rr,\sigma}+H.c.\Big]+U\sum_{\rr}n_{\rr,b}n_{\rr,f}\nonumber\\
  &+V_b\sum_{\langle\rr,\rr'\rangle}n_{\rr',b}n_{\rr,b}+V_f\sum_{\langle\rr,\rr'\rangle}n_{\rr',f}n_{\rr,f},\label{cbl}
\end{align}
and the Haldane-honeycomb (HC) lattice
\begin{align}
  H_{HC}=&\!\sum_{\sigma=b,f}\!\Big[-t\!\!\sum_{\langle\rr,\rr'\rangle}\!\! c_{\rr',\sigma}^{\dag}c_{\rr,\sigma}-t'\!\!\sum_{\langle\langle\rr,\rr'\rangle\rangle}\!\!e^{i\phi_{\rr'\rr}}c_{\rr',\sigma}^{\dag}c_{\rr,\sigma}\nonumber\\
  &-t''\!\!\sum_{\langle\langle\langle\rr,\rr'\rangle\rangle\rangle}\!\!\!\! c_{\rr',\sigma}^{\dag}c_{\rr,\sigma}+H.c.\Big]+U\sum_{\rr}n_{\rr,b}n_{\rr,f}\nonumber\\
  &+V_b\sum_{\langle\rr,\rr'\rangle}n_{\rr',b}n_{\rr,b}+V_f\sum_{\langle\rr,\rr'\rangle}n_{\rr',f}n_{\rr,f},\label{hcl}
\end{align}
where $\sigma=b$ for hardcore bosons and $\sigma=f$ for spinless fermions, $c_{\rr,\sigma}^{\dag}$ ($c_{\rr,\sigma}$) creates (annihilates) a $\sigma$-component particle at site $\rr$, $n_{\rr,\sigma}=c_{\rr,\sigma}^{\dag}c_{\rr,\sigma}$ is the particle number at site $\rr$, $U$ is the strength of the onsite Bose-Fermi repulsion, $V_{\sigma}$ is the strength of the nearest-neighbor repulsion, and $\langle\ldots\rangle$,$\langle\langle\ldots\rangle\rangle$ and $\langle\langle\langle\ldots\rangle\rangle\rangle$ denote the nearest-neighbor, next-nearest-neighbor, and next-next-nearest-neighbor pairs of sites, with the accompanying tunnel couplings $t'=0.3t,t''=-0.2t,\phi=\pi/4$ for the checkerboard lattice, and $t'=0.6t,t''=-0.58t,\phi=2\pi/5$ for the honeycomb lattice respectively (see Figure 1 in Ref.~\cite{Wang2011} for topological lattice geometry).

\textit{Fractional Quantum Hall Effect.---}
In this section, we perform both ED and DMRG calculations on the many-body ground state of the model Hamiltonian Eqs.~\ref{cbl} and~\ref{hcl}, and systematically present results for the topological order at fillings $\nu_b=2/5,\nu_f=1/5$ under strong interactions $U\gg t,V_b=0,V_f\gg t$. Here, we shall elucidate the characteristic $\mathbf{K}=\begin{pmatrix}
2 & 1\\
1 & 3\\
\end{pmatrix}$ matrix from topological degeneracy, topologically invariant Chern number, fractional charge pumping, and entanglement spectrum of the ground states.

\begin{figure}[t]
  \includegraphics[height=1.75in,width=3.4in]{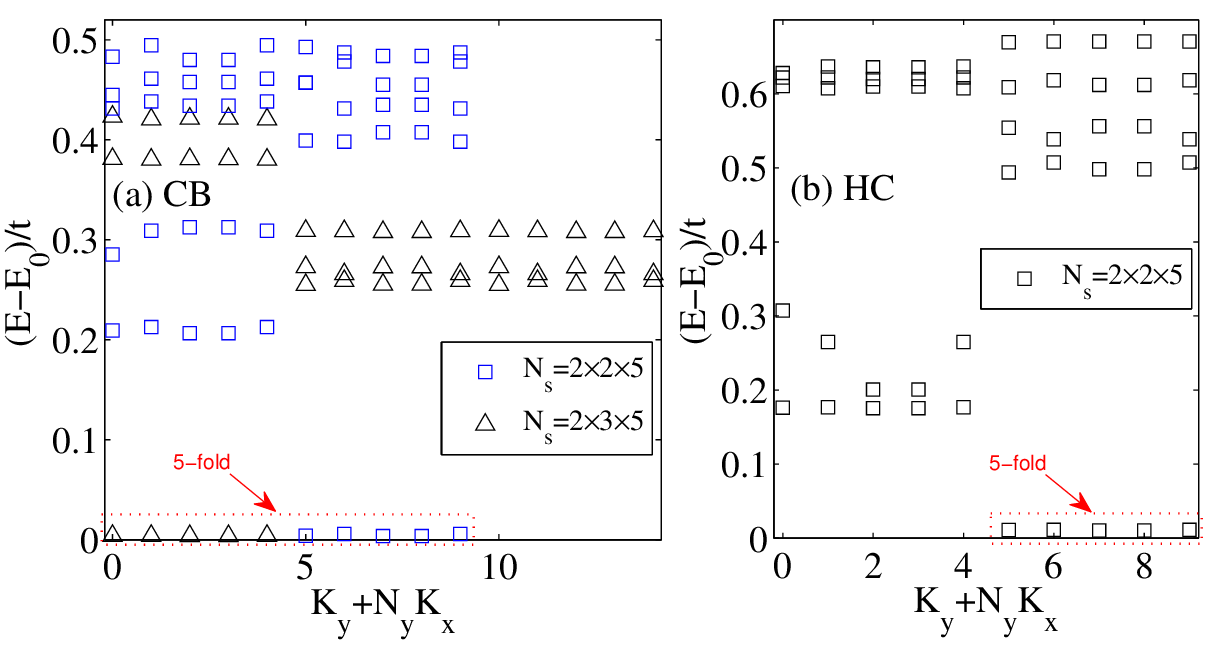}
  \caption{\label{energy} (Color online) Numerical ED results for the low energy spectrum of two-component Bose-Fermi mixtures at $\nu_b=2/5,\nu_f=1/5$ with $U=\infty,V_b=0,V_f/t=100$ in different topological lattices: (a) $\pi$-flux checkerboard lattice and (b) Haldane-honeycomb lattice.}
\end{figure}

First, we demonstrate the topological ground state degeneracy on periodic lattice using the ED study. In small finite periodic lattice systems of $N_x\times N_y$ unit cells (the total number of sites is $N_s=2\times N_x\times N_y$), the energy states are labeled by the total momentum $K=(K_x,K_y)$ in units of $(2\pi/N_x,2\pi/N_y)$ in the Brillouin zone. The particle fillings of the lowest Chern band are fixed as $\nu_{b}=2N_{b}/N_s,\nu_{f}=2N_{f}/N_s$, where $N_{b},N_{f}$ are the global boson/fermion numbers with $U(1)\times U(1)$-symmetry. As shown in Figs.~\ref{energy}(a) and~\ref{energy}(b) for different topological systems in a strongly interacting regime, we find that, there exists a well-defined five-fold degenerate ground states separated from higher energy levels by a robust gap, consistent with the determinant of the $\mathbf{K}$ matrix.

\begin{figure}[b]
  \centering
  \includegraphics[height=1.85in,width=3.4in]{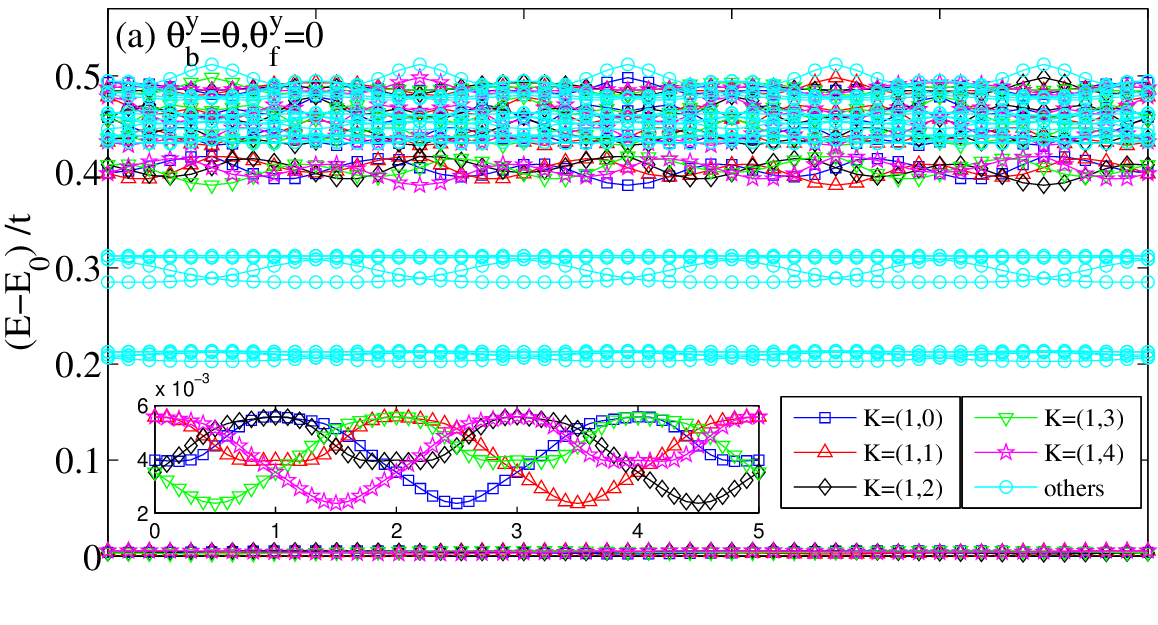}
  \includegraphics[height=1.8in,width=3.4in]{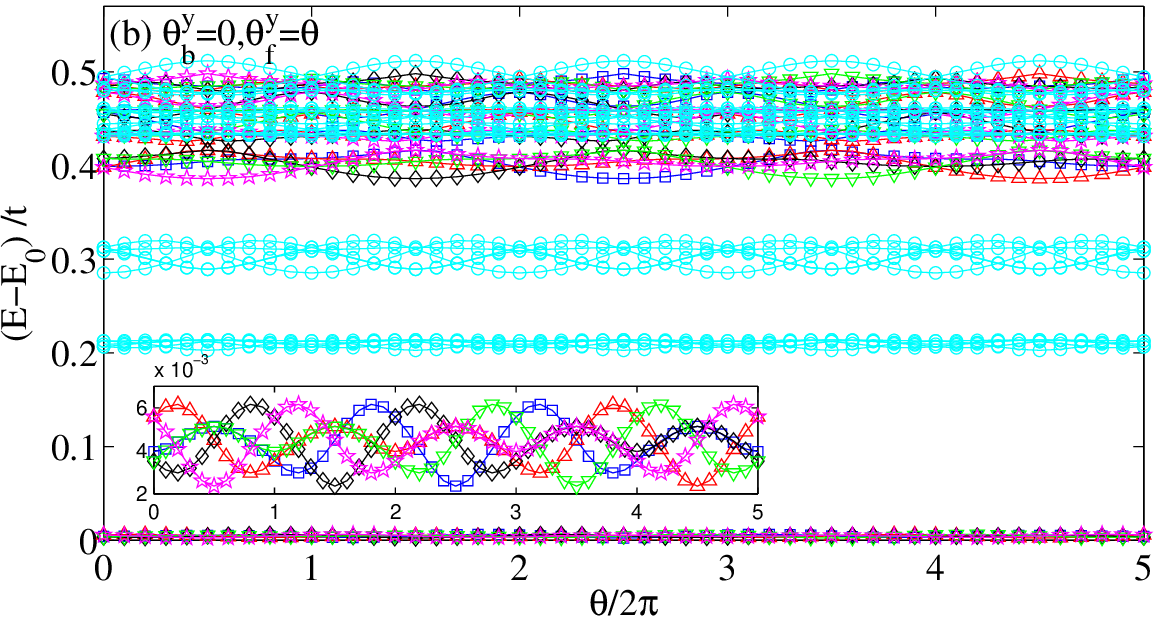}
  \caption{\label{flux} (Color online) Numerical ED results for the low energy spectral flow of Bose-Fermi mixtures $N_{b}=4,N_{f}=2,N_s=2\times2\times5$ with $U=\infty,V_b=0,V_f/t=100$ in topological checkerboard lattice under the insertion of two types of flux quanta: (a) $\theta_{b}^{y}=\theta,\theta_{f}^{y}=0$ and (b) $\theta_{b}^{y}=0,\theta_{f}^{y}=\theta$.}
\end{figure}

To demonstrate the topological equivalence of these ground states,
we further calculate the low energy spectral flux under the insertion of flux quanta $\theta_{\sigma}^{\alpha}$ ($\sigma=b$ for bosons and $\sigma=f$ for fermions) which is the twisted angle of twisted boundary conditions $c_{\rr+N_{\alpha},\sigma}=c_{\rr,\sigma}\exp(i\theta_{\sigma}^{\alpha})$ in the $\alpha=x,y$ direction.
For different types of flux quanta in each species, as shown in Figs.~\ref{flux}(a) and~\ref{flux}(b), we find that these five-fold ground states evolve into each other without mixing with the higher levels,
and the system returns back to itself upon the insertion of five flux quanta for both $\theta_{b}^{\alpha}=\theta,\theta_{f}^{\alpha}=0$
and $\theta_{b}^{\alpha}=0,\theta_{f}^{\alpha}=\theta$.
The robustness of degenerate ground states on different lattice geometries associates the universal internal structures with the behavior of fractional quantization.

\begin{figure}[t]
  \includegraphics[height=2.6in,width=3.4in]{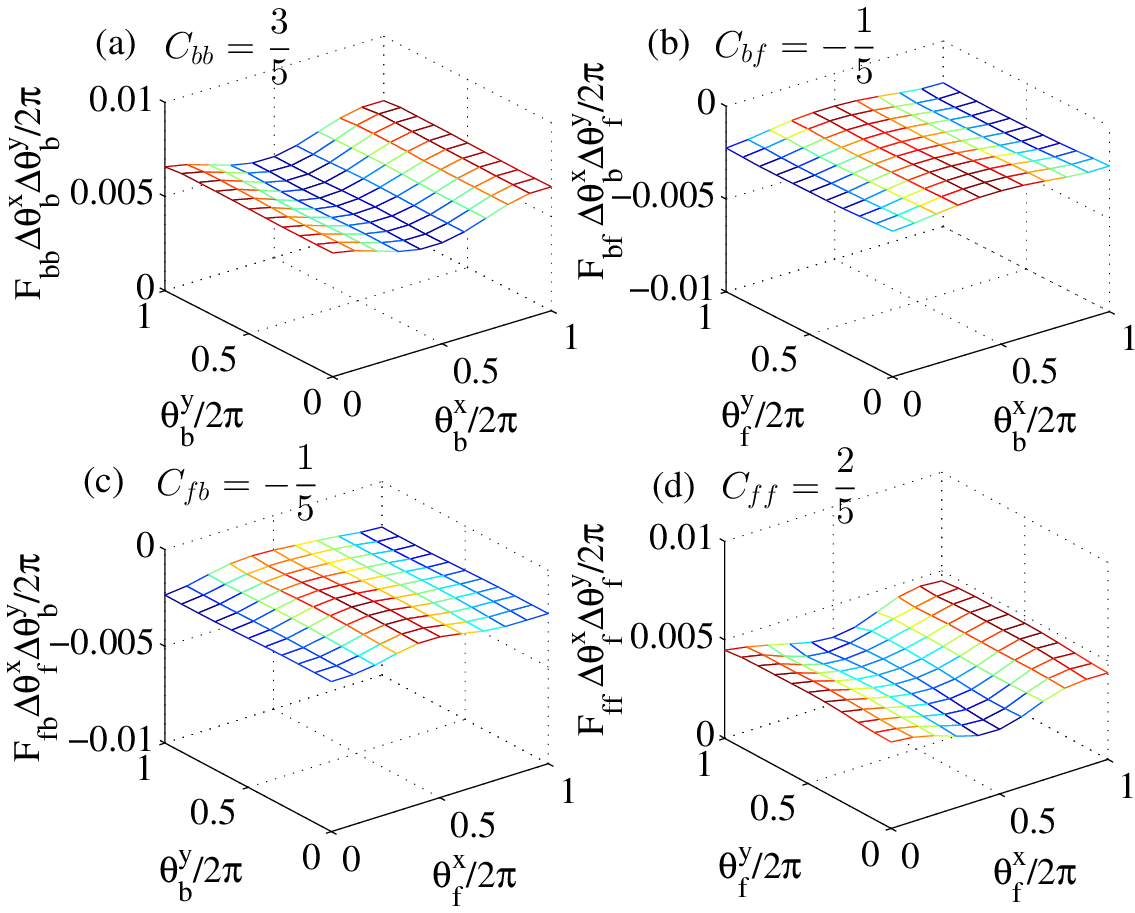}
  \caption{\label{berry} (Color online) Numerical ED results for Berry curvatures $F_{\sigma,\sigma'}^{xy}\Delta\theta_{\sigma}^{x}\Delta\theta_{\sigma'}^{y}/2\pi$ of the ground state at $K=(\pi,0)$ of Bose-Fermi mixtures $N_{b}=4,N_{f}=2,N_s=2\times2\times5$ with $U=\infty,V_b=0,V_f/t=100$ in topological checkerboard lattice under different twisted angles: (a) $(\theta_{b}^{x},\theta_{b}^{y})$, (b) $(\theta_{b}^{x},\theta_{f}^{y})$, (c) $(\theta_{f}^{x},\theta_{b}^{y})$, (d) $(\theta_{f}^{x},\theta_{f}^{y})$.}
\end{figure}

Next we continue to analyze the fractionally quantized topological Hall conductance of the many-body ground state wavefunction $\psi$ for interacting systems~\cite{Niu1985,Sheng2006}, characterized by the Chern number matrix $\mathbf{C}=\begin{pmatrix}
C_{bb} & C_{bf} \\
C_{fb} & C_{ff} \\
\end{pmatrix}$ for Bose-Fermi mixtures. Within the parameter plane of twisted angles $\theta_{\sigma}^{x}\subseteq[0,2\pi],\theta_{\sigma'}^{y}\subseteq[0,2\pi]$, the Chern number of the many-body ground state is defined by $C_{\sigma,\sigma'}=\int\int d\theta_{\sigma}^{x}d\theta_{\sigma'}^{y}F_{\sigma,\sigma'}(\theta_{\sigma}^{x},\theta_{\sigma'}^{y})/2\pi$ with the Berry curvature
\begin{align}
  F_{\sigma,\sigma'}(\theta_{\sigma}^{x},\theta_{\sigma'}^{y})=\mathbf{Im}\left(\langle{\frac{\partial\psi}{\partial\theta_{\sigma}^x}}|{\frac{\partial\psi}{\partial\theta_{\sigma'}^y}}\rangle
-\langle{\frac{\partial\psi}{\partial\theta_{\sigma'}^y}}|{\frac{\partial\psi}{\partial\theta_{\sigma}^x}}\rangle\right).\nonumber
\end{align}
For a given ground state at momentum $K$, by numerically calculating the Berry curvatures using $m\times m$ mesh Wilson loop plaquette in the boundary phase space with $m\geq10$, we obtain the quantized topological invariant $C_{\sigma,\sigma'}$.  In the ED study of finite system sizes, as indicated in Fig.~\ref{berry},
we obtain that the ground state at $K=(\pi,0)$ hosts the fractionally quantized Chern numbers, namely the inverse of the $\mathbf{K}$ matrix,
\begin{align}
  \mathbf{C}=\begin{pmatrix}
C_{bb} & C_{bf} \\
C_{fb} & C_{ff} \\
\end{pmatrix}=\frac{1}{5}\begin{pmatrix}
3 & -1 \\
-1 & 2 \\
\end{pmatrix}=\mathbf{K}^{-1}.\label{chern}
\end{align}
In Chern-Simons gauge field theory, the gauge fields $\alpha_{I}$ (index $I=b,f$) are mutually coupled through the Lagrangian $\sum_{I,J}K_{I,J}\alpha_{I}\varepsilon\partial\alpha_{J}/4\pi+\sum_I\alpha_Ij_I$~\cite{Wen1992a}, and once these gauge fields are integrated out, the left part is given by the mutual coupling current $\sum_{I,J}\pi K^{-1}_{I,J}j_{I}\varepsilon\partial/(\partial)^2j_{J}$. This leads to the Hall current response in Eq.~\ref{chern} when $\theta_{b}^{\alpha},\theta_{f}^{\alpha}$ are also regarded as U(1) gauge fields acting on particles.

\begin{figure}[t]
  \includegraphics[height=1.8in,width=3.4in]{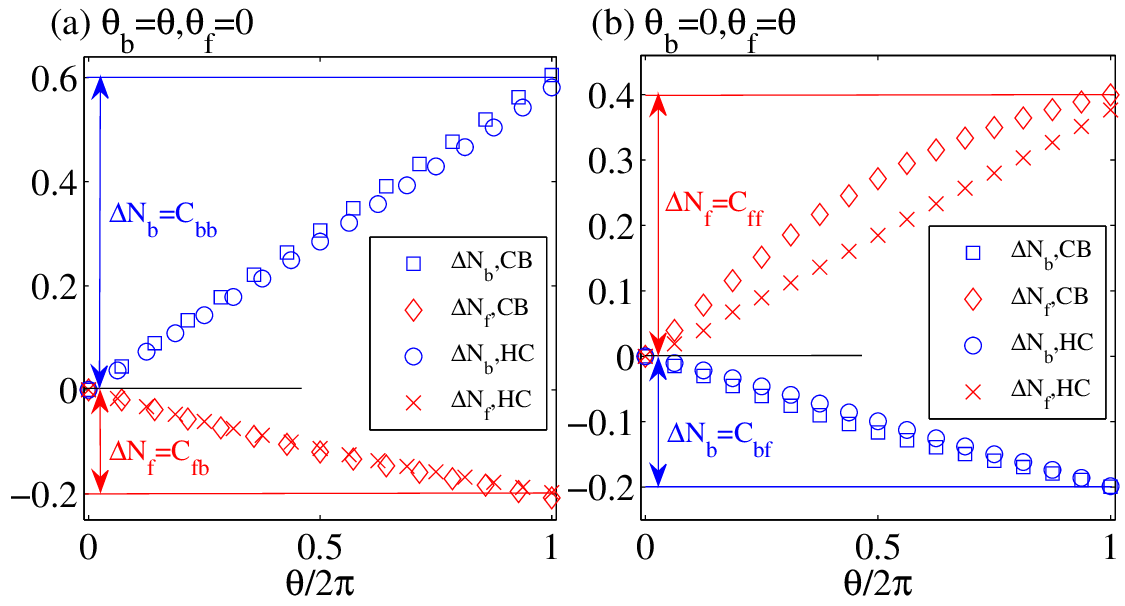}
  \caption{\label{pump} (Color online) Fractional charge transfers for Bose-Fermi mixtures at $\nu_b=2/5,\nu_f=1/5$ with with $U=\infty,V_b=0,V_f/t=100$ on the cylinder of different topological lattices under two types of the insertions of flux quantum: (a) $\theta_{b}^{y}=\theta,\theta_{f}^{y}=0$ and (b) $\theta_{b}^{y}=0,\theta_{f}^{y}=\theta$. Here finite DMRG is used with cylinder width $N_y=3$ and cylinder length $N_x=30$.}
\end{figure}

For larger system sizes, following Laughlin's arguments, we perform the adiabatic thread of one flux quantum on cylinder systems through both the finite and infinite DMRG to calculate the charge pumping, in connection with the quantized Hall conductance~\cite{Laughlin1981,Gong2014}. For multicomponent FQH effect, it is shown in Refs.~\cite{Zeng2017,Zeng2018} that distinct quantized charge responses for each component will be visualized from the right side to the left side upon inserting one flux quantum $\theta_{\sigma}^{y}=\theta,\theta_{\sigma'}^{y}=0$ in only one component from $\theta=0$ to $\theta=2\pi $. In the DMRG, the geometry of cylinders is open boundary condition in the $x$ direction and periodic boundary condition in the $y$ direction. Numerically we cut the cylinder into left-half and right-half parts along the $x$ direction,
and the net charge transfers for each component from the right side to the left side on the cylinder are encoded by $N_{\sigma}(\theta_{\sigma'}^{y})=tr[\widehat{\rho}_L(\theta_{\sigma'}^{y})\widehat{N}_{\sigma}]$
as a function of $\theta_{\sigma'}^{y}$ (here $\widehat{\rho}_L$ the reduced density matrix of the left part, classified by the quantum numbers $\Delta Q_{b},\Delta Q_{f}$). As illustrated in Figs.~\ref{pump}(a) and~\ref{pump}(b) for different system sizes in different topological lattices, we find the universal conclusions that (i) for the flux thread $\theta_b=\theta,\theta_f=0,\theta\subseteq[0,2\pi]$, the charge pumpings
\begin{align}
  &\Delta N_{b}=N_{b}(2\pi)-N_{b}(0)\simeq C_{bb}=\frac{3}{5}, \nonumber\\
  &\Delta N_{f}=N_{f}(2\pi)-N_{f}(0)\simeq C_{fb}=-\frac{1}{5} \nonumber
\end{align}
and (ii) for the flux thread $\theta_b=0,\theta_f=\theta,\theta\subseteq[0,2\pi]$, the charge pumpings
\begin{align}
  &\Delta N_{b}=N_{b}(2\pi)-N_{b}(0)\simeq C_{bf}=-\frac{1}{5},\nonumber\\
  &\Delta N_{f}=N_{f}(2\pi)-N_{f}(0)\simeq C_{ff}=\frac{2}{5} \nonumber
\end{align}
in agreement with the Chern number matrix in Eq.~\ref{chern}.

\begin{figure}[t]
  \includegraphics[height=1.9in,width=3.4in]{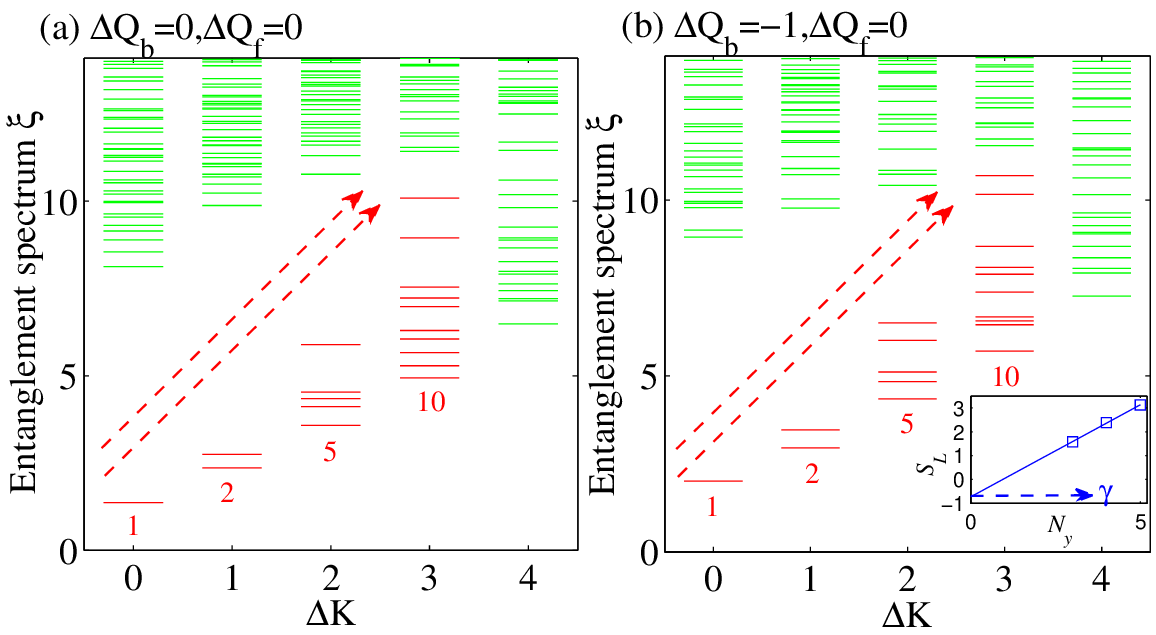}
  \caption{\label{es2} (Color online) Chiral edge mode identified from the momentum-resolved entanglement spectrum for Bose-Fermi mixtures at $\nu_b=2/5,\nu_f=1/5$ with with $U=\infty,V_b=0,V_f/t=100$ on the infinite $N_y=5$ cylinder of topological honeycomb lattice. The horizontal axis shows the relative momentum $\Delta K=K_y-K_{y}^{0}$ (in units of $2\pi/N_y$). Numbers below the dashed red line label the level counting $1,2,5,10$ at different momenta for typical charge sectors (a) $\Delta Q_b=0,\Delta Q_f=0$ and (b) $\Delta Q_b=-1,\Delta Q_f=0$. The inset depicts the DMRG extraction of topological entanglement entropy $\gamma$ in the area law $S_L(N_y)=\alpha N_y-\gamma$ with the maximal bond dimension up to $8500$. }
\end{figure}

In addition, as a hallmark signature of multicomponent FQH effect, the characteristic chirality of edge modes and the concomitant level counting revealed through a low-lying entanglement spectrum in the bulk~\cite{Li2008}, also play one of the key roles in identifying the topological order. For Abelian FQH effect of Bose-Fermi mixtures, the bosonic collective edge modes (so called Luttinger liquids) should be determined by the signs of the eigenvalues of the $\mathbf{K}$ matrix~\cite{Wen1992edge,Wen1995}. Here we examine the structure of the momentum-resolved entanglement spectrum for different cylinder widths $N_y$. As shown in Fig.~\ref{es2}, on the $N_y=5$ cylinder, we observe two forward-moving branches of the low-lying bulk entanglement spectrum with the level counting $1,2,5,10$ for different charge sectors. Nevertheless, this universal level counting is consistent with SU$(3)$ Wess-Zumino-Witten conformal field theory at level one, implying the gapless nature of two gapless free bosonic edge theories. Meanwhile, we also extract the topological entanglement entropy $\gamma$ via the area law of half-cylinder entanglement entropy $S_L(N_y)=\alpha N_y-\gamma$, which encodes information about quantum dimensions of anyons~\cite{Kitaev2006,Levin2006}. For the Abelian FQH effect $\gamma=\ln\sqrt{\det\mathbf{K}}$. The linear fit in the inset of Fig.~\ref{es2}(b) indeed give a topological entanglement entropy $\gamma\sim0.73$ which is remarkably close to $\ln\sqrt{5}$.

Finally, in terms of composite fermion theory, the FQH effect of Bose-Fermi mixtures classified by the $\mathbf{K}=\begin{pmatrix}
m & 1\\
1 & n\\
\end{pmatrix}$ matrix can be understood as follows. Each boson is bound to $(m-1)$ intracomponent flux quanta (or vortices) and one intercomponent flux quantum, while each fermion carries $(n-1)$ intracomponent flux quanta and one intercomponent flux quantum, forming the so-called two-component composite fermions~\cite{Jain2019}. These two-component composite fermions fully occupy the remanent unattached flux quanta in each component, interpreted as integer quantum Hall effect of composite fermion. In flat band models, the presence of a strong nearest-neighbor repulsion between the same component particles is shown to bring about the Laughlin $\nu=1/4$ FQH effect with three flux quanta attached to each hardcore boson~\cite{Wang2011}, and the Laughlin $\nu=1/3$ FQH effect with two flux quanta attached to spinless fermion~\cite{Sheng2011}, while one intercomponent flux quantum is captured by each particle in the opposite component in the presence of the onsite intercomponent repulsion, such as Halperin $(m,m,1)$ FQH effects~\cite{Zeng2017}. Thus when the strong nearest-neighbor Bose-Bose repulsion $V_b\gg t$ is introduced, in addition to strong Bose-Fermi and Fermi-Fermi repulsions $U,V_f\gg t$, we tentatively propose the $\mathbf{K}=\begin{pmatrix}
4 & 1\\
1 & 3\\
\end{pmatrix}$ matrix for possible FQH effect of Bose-Fermi mixtures at fillings $\nu_b=2/11,\nu_f=3/11$. In analogy to the FQH effect at $\nu_b=2/5,\nu_f=1/5$, it is expected to apply the charge pumping under the insertion of flux quantum on the infinite cylinder to identify the fractionally quantized Chern number matrix. However due to the huge wavefunction unit cell (an integer multiple of 22 lattice sites) at low commensurate fillings $\nu_b=2/11,\nu_f=3/11$, we experience a strong numerical difficulty in getting DMRG simulation converged, and have not confirmed the exact fractionally quantized pumping, where a solid investigation of its $\mathbf{K}$ matrix requires a more powerful computational approach beyond the scope of this work.

\textit{Conclusion and outlook.---}
To summarize, we introduced topological lattice models of interacting Bose-Fermi mixtures, and numerically demonstrated the emergence of fractional quantum Hall effects of Bose-Fermi mixtures emanating from the interplay of Bose-Bose, Fermi-Fermi and Bose-Fermi repulsions. Our complementary ED and DMRG simulations further reveal the $\mathbf{K}=\begin{pmatrix}
2 & 1\\
1 & 3\\
\end{pmatrix}$ matrix classification at boson filling $\nu_b=2/5$ and fermion filling $\nu_f=1/5$ with three pieces of major evidence: (i) topological $\det\mathbf{K}$-fold ground-state degeneracies, (ii) fractionally quantized topological Chern number matrix $\mathbf{C}=\mathbf{K}^{-1}$, and (iii) two parallel-propagating chiral edge branches according to positive signs of two eigenvalues of the $\mathbf{K}$ matrix. Also we provide a physical mechanism based on two-component composite fermion, and suggest the possible emergence of FQH effect at boson filling $\nu_b=2/11$ and fermion filling $\nu=3/11$ classified by the $\mathbf{K}=\begin{pmatrix}
4 & 1\\
1 & 3\\
\end{pmatrix}$ matrix when a strong nearest-neighbor intracomponent repulsion is introduced. Finally, in consideration of short-range interactions in cold atom physics, our flat band models of Bose-Fermi mixtures contribute as the ubiquitous scenarios of a microscopic Hamiltonian featuring FQH effects, and may provide a promising perspective on the exploration of new phases of matter using ultracold atomic mixtures in future experimental studies of topological bands~\cite{Cooper2019}.

\textit{Acknowledgements.---}
T.-S.Z. thanks D. N. Sheng and W. Zhu for inspiring discussions and prior collaborations on multicomponent fractional quantum Hall physics in topological flat band models.
This work is supported by the National Natural Science Foundation of China (NSFC) under Grant No. 12074320.

\end{document}